\documentclass[apj,square,numbers]{emulateapj}
\usepackage{graphicx}
\usepackage{xcolor}
\usepackage{amsmath,amssymb,latexsym,bm}
\usepackage{hyperref}
\usepackage{ulem}
\begin{document}
\title{Evolutions and Calibrations of Long Gamma-Ray Bursts Luminosity Correlations Revisited}

\author{Guo-Jian Wang$^{1}$, Hai Yu$^{2,3,1}$, Zheng-Xiang Li$^{1}$, Jun-Qing Xia$^{1,\ast}$, Zong-Hong Zhu$^{1}$}

\affil{$^{1}$Department of Astronomy, Beijing Normal University, Beijing 100875, China; xiajq@bnu.edu.cn}
\affil{$^{2}$School of Astronomy and Space Science, Nanjing University, Nanjing 210093, China}
\affil{$^{3}$Key Laboratory of Modern Astronomy and Astrophysics (Nanjing University), Nanjing 210093, China}

\begin{abstract}

Luminosity correlations of long Gamma-ray bursts (GRB) are extensively proposed as an effective complementarity to trace the Hubble diagram of Universe at high redshifts, which is of great importance to explore properties of dark energy. Recently, several empirical luminosity correlations have been statistically proposed from GRB observations. However, to treat GRB as the distance indicator, there are two key issues: the redshift evolution of luminosity correlations and their calibrations. In this paper, we choose the Amati relation, the correlation between the peak spectra energy and the equivalent isotropic energy of GRBs ($E_{\rm p}-E_{\rm iso}$), as an example, and find that the current GRB dataset implies that there could be a evolution of the luminosity correlation with respect to the redshift. Therefore, we propose an extended Amati relation with two extra redshift-dependent terms to correct the redshift evolution of GRB relation. Secondly, we carefully check the reliability of the calibration method using the low-redshift GRB data. Importantly, we find that the low-redshift calibration method does not take whole correlations between $\Omega_{\rm m}$ and coefficients into account. Neglecting these correlation information can break the degeneracies and obtain the biased constraint on $\Omega_{\rm m}$ which is very sensitive to values of parameters for the calibration. A small shift of parameters of ``calibrated" relation could significantly change the final constraint on $\Omega_{\rm m}$ in the low-redshift calibration method. Finally, we simulate several GRB samples with different statistical errors and find that, in order to correctly recover the fiducial value of $\Omega_{\rm m}$ using the low-redshift calibration method, we need a large number of GRB samples with high precisions.

\end{abstract}


\maketitle

\section{Introduction}\label{sec:section1}
Since the accelerating expansion of the universe was revealed by observations of type Ia supernovae (SN) ~\citep{Riess:1998, Perlmutter:1999}, SN plays an essential role to study the expansion history of the Universe and the nature of dark energy. However, due to the limited intrinsic luminosity and the extinction from the interstellar medium, the maximum redshift of the SN we can currently observe is about $1.7$. Approximately, for most other popular probes at low redshift, such as baryon acoustic oscillations (BAO) and weak lensing, the bottleneck of redshift range in the near future observations is about $z\sim2$. Whereas fluctuations of the cosmic microwave background (CMB) provide cosmological information at the last scattering surface ($z\sim1089$). Therefore, we do not have many effective methods to observe the evolution of our Universe at $2<z<1089$, usually referred as the ``cosmological desert''.

Gamma-ray bursts (GRB) are the most energetic explosions after the big bang in the Universe. The equivalent isotropic energy radiated in a few seconds can reach up to $10^{48}\sim10^{53}$ erg (see e.g.,~\citep{Piran:1999, Meszaros:2002, Meszaros:2006, Kumar and Zhang:2015} for recent reviews). Thanks to the extreme brightness and the immunity to dust extinction of high-energy photons in the gamma-ray band, GRB are detectable up to redshift $z\sim 10$ ~\citep{Salvaterra:2009, Tanvir:2009, Cucchiara:2011, Tanvir:2013}. Therefore, in the literatures, GRB are widely proposed, as a complementary data of SN, to trace the Hubble diagram of the Universe at high redshifts. However, different from the consistent luminosities of SN which allows us to qualify SN as an ideal distance indicator, the central engine mechanism of explosions of GRB has not yet clearly been understood. Therefore, drawing cosmological implications from GRB observations is quite intractable. Recently, many methods have been proposed to achieve some progresses, in order to treat GRB as the distance indicator ~\citep{Schaefer:2002, Dai:2004tq, Ghirlanda:2004a, Liang:2005, Firmani:2005, Schaefer:2007, Liang:2008, Li:2008, Wei:2009, Wei:2010, Liu:2015, Li:2015, Lin:2016,Wang:2015}.

Luminosity or energy correlations of GRB are empirical connections between measurable properties of the prompt gamma-ray emission and the luminosity or energy of GRB. In recent years, several empirical luminosity correlations have been statistically proposed from observations, such as various popular two-variable relations: the correlation between spectrum lag and isotropic peak luminosity ($\tau_{\rm lag}-L$)~\citep{Norris:2000}; the correlation between time variability and isotropic peak luminosity ($V-L$)~\citep{Fenimore:2000}; the correlation between the peak energy of $\nu F_{\nu}$ spectrum and isotropic equivalent energy ($E_{\rm p}-E_{\rm iso}$, the Amati relation)~\citep{Amati:2002}; the correlation between peak energy and collimation-corrected energy ($E_{\rm p}-E_{\gamma}$)~\citep{Ghirlanda:2004b}; the correlation between peak energy and isotropic peak luminosity ($E_{\rm p}-L$)~\citep{Schaefer:2002,Yonetoku:2004}; the correlation between minimum rise time of light curve and isotropic peak luminosity ($\tau_{\rm RT}-L$)~\citep{Schaefer:2007}; and some well-known multi-variable relations, such as the correlation between $E_{\rm iso}$, $E_{\rm p}$, and the break time of the optical afterglow light curves $t_{\rm b}$~\citep{Liang:2005}. The general expression for luminosity correlations is
\begin{equation}\label{eq1}
  y=a+bx~,
\end{equation}
where $x=\log E_{\rm p}$ (or $\log V$, $\log\tau_{\rm lag}$, $\log\tau_{\rm RT}$), $y=\log L$ (or $\log E_{\rm iso}$, $\log E_{\gamma}$). Determinations of coefficients $a$ and $b$ of correlations could tell us whether these empirical luminosity correlations are independent on the redshift, which is very crucial for treating GRB as the distance indicator to investigate the evolution of our Universe. In addition, some new models for GRB have been proposed in resent years, such as the spectro-temporal multi-component model \citep{Guiriec:2011,Guiriec:2013,Guiriec:2015a,Guiriec:2015b,Guiriec:2016a,Guiriec:2016b}. Interestingly, this model could result in a new time-resolved relation: the relation between the luminosity of the non-thermal component, $L^{\rm nTH}_{\rm i}$, and its corresponding $\nu F_{\nu}$ spectral peak energy in the rest frame, $E^{\rm NT,rest}_{\rm peak,i}$ (i.e., $L^{\rm nTH}_{\rm i}-E^{\rm NT,rest}_{\rm peak,i}$ relation).

More importantly, determinations of coefficients can also show the reliability of the calibration of GRB luminosity correlations directly. In the literatures, there are two methods to calibrate the luminosity correlations: ``self-calibration'' method, which is based on the global fitting technique using all GRB data to constrain the cosmological parameters and the coefficients of luminosity correlations simultaneously~\citep{Schaefer:2002, Li:2008}; ``low-redshift calibration'' method, which is using the low-redshift GRB data to obtain the constraints on coefficients of luminosity correlations, and directly extrapolate them to high redshifts to constrain the cosmological parameters using the high-redshift GRB data~\citep{Liang:2008, Wei:2009, Wei:2010, Ding:2015, Liu:2015, Lin:2016, Wang:2016}.

Apparently, these two methods are seriously dependent on the assumption that empirical luminosity correlations are universal and do not evolve with respect to redshift. In practice, for the Amati relation, \citet{Li:2007} first revealed that the systematically significant variations of the intercept $a$ and the slope $b$ with the cosmological redshift. Later, \citet{Basilakos:2008} investigated the same issue for several other luminosity relations. Due to the limited quality of GRB samples, there was no statistically significant evidence for evolution of the calibration parameters. In addition, as a cosmological probe, the evolution of GRB luminosity correlations may result in overestimate or underestimate of cosmological parameters~\citep{Dainotti:2013}. More recently, in~\citet{Lin:2015}, they re-investigated the Amati relation using low-redshift ($z < 1.4$) and high-redshift ($z > 1.4$) GRB, and found that the coefficients of Amati relation in low-redshift GRB differs from those of high-redshift GRB at more than $3\sigma$ confidence level. Therefore, they finally concluded that long gamma-ray bursts might not be standard candles. Following on this topic, \citet{Lin:2016} further examined the possible redshift dependence of several other luminosity correlations and found that only for the $E_{\rm p} - E_\gamma$ relation, the low-redshift GRB could give the similar constraints on the coefficients with those from the high-redshift GRB within $1\sigma$ confidence level.

In this paper, we choose the Amati relation as an example to investigate the evolutions and calibrations of GRB luminosity correlation with the latest GRB observational data~\citep{Liu:2015}. Firstly, we divide the whole GRB sample into two bins ($z<1.4$ and $z>1.4$). In each bin, we use the GRB data to constrain the coefficients in the $\Lambda$CDM framework to check whether they are independent on the redshift. Then, we propose an extended Amati relation by introducing two extra terms to characterize the evolutions with respect to the redshift. Secondly, we carefully check the reliability of the calibration method. In practice, we use the low-redshift GRB data ($z<1.4$) to calibrate the coefficients of GRB luminosity correlation and directly extrapolate them to high-redshift events. Then we use this ``calibrated'' Amati relation to constrain the matter density parameter $\Omega_{\rm m}$ with the high-redshift GRB data. Finally, we simulate several samples of GRB data with high precisions to investigate that in which conditions the GRB data calibrated by the low-redshift calibration method can be used for cosmological studies.

This paper is organized as follows: Section \ref{sec:evolution} and Section \ref{sec:calibration} are dedicated to present the method and numerical results on the analysis of evolution and calibration of GRB luminosity correlation. Finally, conclusions and discussions are presented in section \ref{sec:summary}.

\section{Evolutions of Luminosity Correlations}\label{sec:evolution}

\subsection{Amati Relation \& GRB Data}\label{sec:section2.1}

Although for recent years, many empirical luminosity correlations have been statistically concluded from long GRB observations, the Amati correlation is the most widely used one among them for cosmological studies. The original version of the Amati relation is expressed as
\begin{align}
\log{\frac{E_{\text{iso}}}{\text{erg}}}&=a+b\log{\frac{E_{{\rm p},i}}{300~\text{keV}}}~,\label{equ:relation}
\end{align}
where $E_{{\rm p},i}=E_{\rm p}(1+z)$ to correct the redshift dilation of the spectrum~\citep{Wang:2011}. The isotropic equivalent energy $E_{\rm iso}$ can be calculated from the bolometric fluences $S_{\rm bolo}$~\citep{Schaefer:2007}:
\begin{equation}\label{equ:Eiso}
E_{\rm iso}=4\pi d^{2}_{L}S_{\rm bolo}\left(1+z \right)^{-1}~.
\end{equation}
The uncertainty of $E_{\rm iso}$ propagates from the uncertainties of $S_{\rm bolo}$ and $d_{\rm L}$. $S_{\rm bolo}$ is calculated from the observed peak photon flux in the rest frame $1-10,000$ keV energy band by assuming the Band spectrum~\citep{Band:1993}. To calculate the luminosity distance $d_{\rm L}$, we assume a flat concordance $\Lambda$CDM model with the Hubble constant: $H_0=67.8~\rm km~s^{-1}Mpc$ and the matter density parameter: $\Omega_{\rm m}=0.308$ obtained from the latest $Planck$ 2015 results~\citep{Ade:2015}. Hence, $d_{\rm L}$ in Eq.~(\ref{equ:Eiso}) can be written as
\begin{equation}\label{equ:luminosity}
d_{\rm L}(z)=\frac{c(1+z)}{H_0}\int^{z}_{0}\frac{dz}{\sqrt{\Omega_{\rm m}(1+z)^3+(1-\Omega_{\rm m})}}~.
\end{equation}

\begin{table*}
\centering
\caption{$1\,\sigma$ Constraints on the intrinsic scatters ($\sigma_{\rm int}$) and luminosity correlation parameters ($a$, $b$, $\alpha$ and $\beta$) in the flat $\Lambda$CDM framework for the original Amati relation and extended Amati relation respectively. For comparison, we also show the constraints using the global fitting method.}\label{tab:parameters_LCDM}
\begin{tabular}{llccccccc}
\hline\hline
relation   & bins & $\sigma_{\rm int}$ & N & $\chi^2_{\rm min}$ & $a$ & $b$ & $\alpha$ & $\beta$\\
\hline
&low-$z$& $0.40\pm0.05$ & $59$ & $55.51$ & $52.75\pm0.06$ & $1.60\pm0.10$ & &\\
Amati relation:&high-$z$& $0.32\pm0.04$ & $79$ & $76.18$ & $52.95\pm0.05$ & $1.30\pm0.12$ & &\\
&Global Fit& $0.35\pm0.03$ & $138$ & $134.83$ & $52.62\pm0.07$ & $1.49\pm0.07$ & &\\
\hline
Extended &low-$z$& $0.40\pm0.05$ & $59$ & $55.15$ & $52.42\pm0.35$ & $1.46\pm0.29$ & $0.72\pm0.75$ & $0.14\pm0.64$\\
Amati relation: &high-$z$& $0.32\pm0.04$ & $79$ & $74.06$ & $53.42\pm0.46$ & $0.08\pm1.01$ & $-0.67\pm0.64$ & $1.73\pm1.42$\\
&Global Fit& $0.35\pm 0.03$ & $138$ & $134.65$ & $52.52\pm0.16$ & $1.59\pm0.18$ & $0.33\pm 0.40$ & $-0.31\pm0.32$\\
\hline\hline
\end{tabular}
\end{table*}

\begin{figure}[t]
	\centering
	\includegraphics[width=0.45\textwidth]{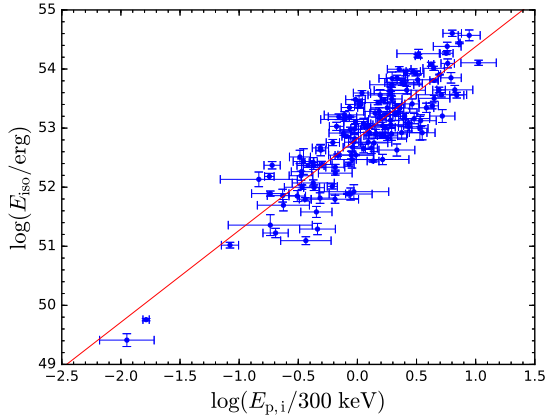}
	\caption{The original Amati relation for all GRB data. The blue points with error bars are GRB data. The red line denotes the theoretical prediction from the best-fit values $a=52.82$ and $b=1.56$, which are derived from maximizing the D'Agostini's likelihood.}\label{fig:Amati_relation}
\end{figure}

\begin{figure}[t]
	\centering
	\includegraphics[width=0.43\textwidth]{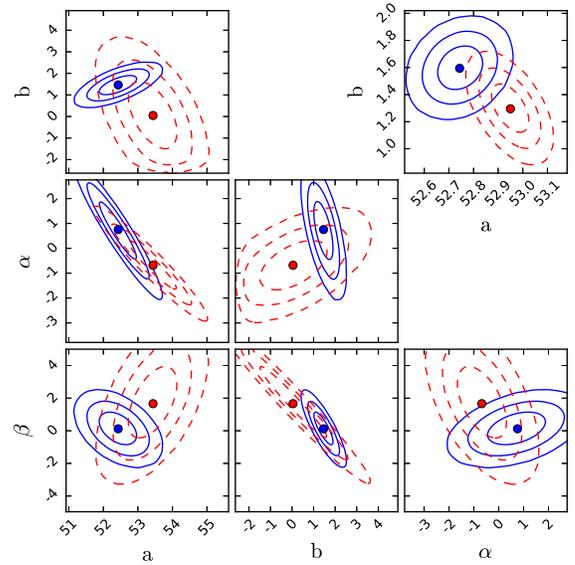}
	\caption{The 2-dimensional marginalized distributions with 1$\sigma$, 2$\sigma$, and 3$\sigma$ contours for the coefficients $a$, $b$, $\alpha$, and $\beta$ for the original Amati relation (the upper right panel) and the extended Amati relation (other six panels), respectively. The blue and red contours denote the results on the parameters of the GRB relations from the low-redshift and high-redshift GRB subsamples, respectively. The central points represent the best-fit values.}\label{fig:testnoevo}
\end{figure}

To test the possible redshift dependence of the Amati luminosity correlation, we use the GRB sample compiled in~\citet{Liu:2015}, which includes 138 well-measured GRB in the redshift range: $z\in [0.0331,~8.1]$. In general, we examine the evolution of the coefficients $a$ and $b$ according the following steps: Firstly, we get the equivalent isotropic energy $E_{\rm iso}$ for each GRB at redshift $z$ with the distance in Eq.~(\ref{equ:luminosity}) taken into account. Then, we divide all observed 138 GRB sample into two redshift bins: [0.0331, 1.4], [1.4, 8.1]. The reason of choosing $z = 1.4$ as the threshold is that the maximal redshift of SNe Ia data samples we are using for cosmological studies is usually around $z=1.4$. The Universe below $z = 1.4$ has already been well studied by using the SNe Ia datasets \citep{Amanullah:2010, Suzuki:2012, Betoule:2014}. Therefore, we treat the GRB data with $z < 1.4$ and $z > 1.4$ as the low-z and high-z samples, respectively. In these two redshift bins, there are 59 and 79 GRB data, respectively. And then, we constrain the luminosity correlation coefficients $a$ and $b$ by maximizing the D'Agostini's likelihood~\citep{DAgostini:2005}:
\begin{align}\label{D_like}
\nonumber \mathcal{L}_D(a,b,\sigma_{\rm int})\propto
&\prod_i\frac{1}{\sqrt{\sigma_{\rm int}^2+\sigma_{y_i}^2+b^2\sigma_{x_i}^2}}\\
&\times \exp\left[-\frac{(y_i-a-bx_i)^2}{2(\sigma_{\rm int}^2+\sigma_{y_i}^2+b^2\sigma_{x_i}^2)}\right],
\end{align}
or, equivalently, by minimizing the $\chi^2$:
\begin{align}
\nonumber\chi^{2}_{D}(a,b,\sigma_{\rm int})=&\sum_{i}\ln \left(\sigma^{2}_{\rm int}+\sigma^{2}_{y_{i}}+b^{2}\sigma_{x_{i}}^{2}\right)\\
&+\sum_{i}\frac{\left(y_{i}-a-bx_{i}\right)^{2}}{\sigma^{2}_{\rm int}+\sigma^{2}_{y_{i}}+b^{2}\sigma_{x_{i}}^{2}},
\end{align}
where $\sigma_{\rm int}$ is the intrinsic scatter which represents any other unknown uncertainties except for the observational statistical ones. Here, in order to test the feasibility of D'Agostini's likelihood, we constrain the two parameters of Amati relation use all GRB data (138 GRB) in the flat $\Lambda$CDM framework firstly. The minimum chi-squared value is $\chi^2_{\rm min}=134.91$ and $\chi^2_{\rm min}/{\rm d.o.f.}=0.98$. We also plot the comparison between the GRB data and the theoretical prediction of the best-fit values in Fig. \ref{fig:Amati_relation}. All these results show that the D'Agostini’s likelihood is feasible to constrain parameters. Finally, after getting constraints on $a$ and $b$, in Fig.\ref{fig:testnoevo} we plot the 2 dimensional marginalized constraints in the plane of ($a$, $b$) from the low-redshift (blue contours) and high-redshift (red contours) GRB data, respectively.

\begin{figure*}[t]
	\centering
	\includegraphics[width=0.8\textwidth]{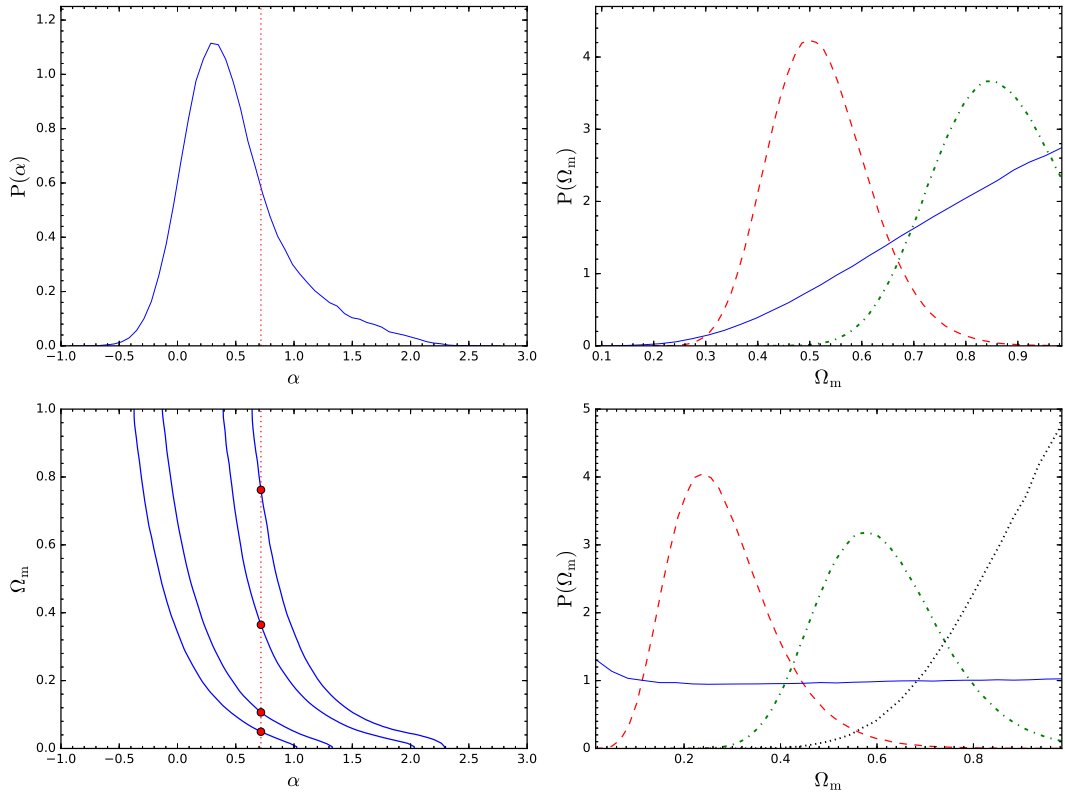}
	\caption{1-dimensional and 2-dimensional marginalized constraints on $\Omega_{\rm m}$ and $\alpha$ for the original Amati relation (the upper right panel) and the extended Amati relation (other three panels). For the extended Amati relation, blue lines of the lower left three panels represent the results from global fitting method. The dotted red lines in the left two panels represent $\alpha=0.72$ (see Table \ref{tab:parameters_LCDM}) which is the best-fit value obtained from the low-redshift GRB data in the low-$z$ calibration method. The constraint on $\Omega_{\rm m}$ from the low-redshift calibration method are shown in the lower right panel (the red dashed line, the green dash dot line, and the black dotted line). For comparison, in the upper right panel we also show the results obtained from the low-redshift calibration method and the global fitting method for the original Amati relation, where the blue line is the result of global fitting method, the red dashed line and green dash dot line are the result from low-redshift calibration method.}\label{fig:Omm}
\end{figure*}

In our analysis, we use {\it emcee}\footnote{https://pypi.python.org/pypi/emcee}~introduced by \citet{Foreman-Mackey:2012}, a Python module that uses the Markov chain Monte Carlo (MCMC) method to get the best-fit values and their uncertainties of parameters $a$, $b$ and $\sigma_{\rm int}$ by generating sample points of the probability distribution. The best-fit values with 1$\sigma$ errors are shown in Table \ref{tab:parameters_LCDM}. We also plot the results in Fig.~\ref{fig:testnoevo} (the upper right panel), in which the blue and red contours denote the results on the parameters of Amati relation from the low-redshift and high-redshift GRB subsamples, respectively. Apparently, these two constraints are not match very well. The best fit values are more than $3\sigma$ away. When considering the uncertainties, the tension between two fitting results is less than $2\sigma$ confidence level, which implies that the parameters of Amati relation might be redshift-dependent. The result is similar with some other works \citep{Lin:2015}.

\subsection{Redshift Evolution of Amati Relation}\label{sec:section2.2}

As the statistically obvious evolution of the luminosity correlation has been revealed, we should be careful to use the GRB data as the distance indicator for cosmological studies. Because all these investigations are based on the assumption that the luminosity correlation does not vary with respect to the redshift, i.e., both $a$ and $b$ in Eq.~(\ref{equ:relation}) should be constant. To solve this problem, inspired by the study for the evolution of light-curve fitting parameters in SN observations~\citep{Wang:2013, Wang:2014} and that for the evolution of the GRB intrinsic luminosity~\citep{Yu:2015}, we introduce two extra redshift-dependent terms to characterize the redshift evolution of the luminosity correlation. Since most redshifts of GRB are much greater than unity, rather than the linear expression in~\citet{Wang:2013, Wang:2014} and the logarithmic parametrization in~\citet{Yu:2015}, we select two mild formulas
\begin{equation}\label{equ2}
  a\rightarrow A=a+\alpha\frac{z}{1+z}~~,~~~b\rightarrow B=b+\beta\frac{z}{1+z}
\end{equation}
to avoid extreme results at high redshifts. Therefore, the extended Amati relation can be expressed as
\begin{align}
\log{\frac{E_{\text{iso}}}{\text{erg}}}&=\bigg(a+\frac{\alpha z}{1+z}\bigg)+\bigg(b+\frac{\beta z }{1+z}\bigg)\log{\frac{E_{\rm p,i}}{300~\text{keV}}},\label{equ:newrelation}
\end{align}
where $a$,$b$,$\alpha$ and $\beta$ are new four coefficients in the relation which should be obtained from the GRB data.

Following the same procedures for analyzing the evolutions of $a$ and $b$ presented in the subsection~\ref{sec:section2.1}, we also investigate the evolutions of luminosity correlation coefficients $a$, $b$, $\alpha$ and $\beta$ using the GRB sample in the same low-redshift and high-redshift subsamples, respectively. The best-fit values with 1$\sigma$ errors are shown in Table \ref{tab:parameters_LCDM}. Fig. \ref{fig:testnoevo} (the lower left six panels) shows the results regarding evolutions of luminosity correlation coefficients ($a$, $b$, $\alpha$ and $\beta$). Since we have two parameters to describe the redshift dependence, compared with the result of original Amati relation, the tension between constraints of parameters from low-$z$ and high-$z$ GRB data are alleviated, most of which are consistent within $1\sigma$ confidence level. Of course, these two extra free parameters could bring the large uncertainties and make the constraints weaker.

\section{Calibrations of Luminosity Correlations}\label{sec:calibration}

\subsection{Low Redshift Calibration}\label{sec:section3.1}

As shown in section \ref{sec:section2.2}, the tension between the constraints on parameters of the original Amati relation from the low-redshift and high-redshift GRB subsamples is more than $1\sigma$ confidence level, while the constraints are consistent within $1\sigma$ confidence level for the extended Amati relation. However, this does not mean that we can safely use the GRB data for cosmological studies. We need to check the calibration of GRB luminosity correlations further. Usually, we have two calibration methods. One is to calibrate luminosity correlations with low-redshift GRB samples in the context of the flat $\Lambda$CDM model and then extrapolate these obtained coefficients to high redshift range. The other method is the global fitting analysis using the MCMC technique in which luminosity correlation parameters and cosmological parameters are simultaneously fitted on the same weight in the context of the flat $\Lambda$CDM model \citep{Li:2008}. Therefore, here we use these two methods to calibrate the extended Amati relation and compare the constraints on the cosmological parameters.

In the low-redshift calibration method, we obtain the luminosity distance of the low-redshift GRB sample ($z < 1.4$) by using Eq.~(\ref{equ:luminosity}). And then we calibrate the luminosity correlation by maximizing the D'Agostini's likelihood (Eq.~\ref{D_like}) for the 59 low-redshift GRBs. The constraints on the luminosity correlation coefficients are shown in Tab. \ref{tab:parameters_LCDM}. Then we directly extrapolate these coefficients to high redshifts. Therefore, we construct the Hubble diagram for the 79 high-redshift GRBs based on the calibrated luminosity correlation and investigate cosmological implication from this Hubble diagram in the context of standard $\Lambda$CDM model. In this framework, the luminosity distance is determined by Eq.~(\ref{equ:luminosity}), and the corresponding distance modulus is
\begin{equation}
\mu(z)=5\lg\frac{d_L(z)}{\rm Mpc}+25.
\end{equation}
We use the numerical method to fit $\Omega_{\rm m}$ and the corresponding $\chi^2$ is
\begin{equation}\label{equ:SNe-glbfit-chi2}
\chi^2 = \sum_i\frac{[\mu_{\rm obs}(z_i)-\mu_{\rm th}(z_i;\Omega_{\rm m})]^2}{\sigma^2_{\mu,i}},
\end{equation}
where $\sigma_{\mu,i}$ is the total uncertainty of distance modulus for the $i$th GRB and it is propagated from the uncertainties of $S_{\rm bolo},~E_{\rm p,i},~E_{\rm iso}$
\begin{equation}\label{error}
\sigma^2_{\mu}=\left(\frac{5}{2}\right)^2\left[\sigma^2_{\log E_{\rm iso}}+\left(\frac{\sigma_{S_{\rm bolo}}}{S_{\rm bolo}\ln 10}\right)^2\right],
\end{equation}
where
\begin{align*}
&\sigma^2_{\log{E_{\rm iso}}}=~\sigma^2_a+\left(\frac{z\sigma_{\alpha}}{1+z}\right)^2+\left(\sigma_{b}\log\frac{E_{\rm p,i}}{300~\rm keV}\right)^2\\
&+\left(\frac{z\sigma_{\beta}}{1+z}\cdot\log\frac{E_{\rm p,i}}{300~\rm keV}\right)^{2}+
\left[\frac{\left(b+\frac{z\beta}{1+z}\right)\sigma_{E_{\rm p,i}}}{E_{\rm p,i}\ln{10}}\right]^2+\sigma^2_{\rm int}.
\end{align*}

Firstly, we consider the original Amati relation. In the upper right panel of Fig. \ref{fig:Omm} (red dashed line), we show the one-dimensional distribution of the matter density parameter $\Omega_{\rm m}$ obtained from the 79 high-redshift GRB data by using the low-redshift calibration method. If we use the best-fit values $a=52.75$ and $b=1.60$ to calibrate the high-redshift GRB data, we obtain the 68\% C.L. constraint on the matter density parameter:
\begin{equation}\label{eq:omm:low-z0}
  \Omega_{\rm m}=0.50\pm0.10~.
\end{equation}
Based on this result, we can see that this obtained constraint on $\Omega_{\rm m}$ is far away from the fiducial value of $\Omega_{\rm m}=0.308$, which is used to constrain the parameter $a$ and $b$ from the low-$z$ GRB data for calibration. The significance of this inconsistence is about $2\sigma$ confidence level.

Then, we use the extended Amati relation to do the calculations, following the same procedures. In the lower right panel of Fig. \ref{fig:Omm} (red dashed line), we also show the one-dimensional distribution of $\Omega_{\rm m}$ obtained from the 79 high-redshift GRB data by using the low-redshift calibration method. The obtained constraints on $\Omega_{\rm m}$ is:
\begin{equation}\label{eq:omm:low-z}
  \Omega_{\rm m}=0.24\pm0.12~,
\end{equation}
at 68\% confidence level, which is consistent with previous works \citep{Wang:2016}. This result is consistent with the fiducial value of the matter density parameter very well, which also implies that the extended Amati relation might be better than the original one.

\begin{figure}
	\centering
	\includegraphics[width=0.45\textwidth]{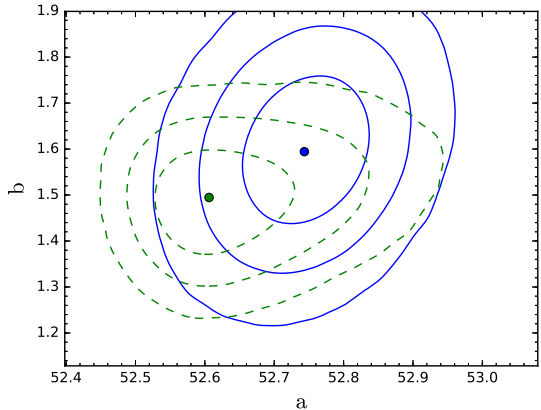}
	\caption{The 2-dimensional marginalized distributions with 1$\sigma$, 2$\sigma$, and 3$\sigma$ contours for the coefficients $a$ and $b$ for the original Amati relation. The blue and green contours denote the results obtained from the low-$z$ GRB data in the low-redshift calibration method and from all GRB data in the global fitting method. The points represent the best-fit values of coefficients.}\label{fig:glbfit_LCDM}
\end{figure}

\subsection{Global Fitting Method}\label{sec:section3.2}

The second method is the global fitting method, in which we calculate the matter density parameter $\Omega_{\rm m}$ and luminosity correlation coefficients ($a$, $b$, $\alpha$ and $\beta$) simultaneously with the same weight. The constraints on these coefficients from all-redshift GRB data are also listed in Table \ref{tab:parameters_LCDM}.

Similar with the discussions in the section \ref{sec:evolution}, we start with the original Amati relation. In Table \ref{tab:parameters_LCDM} and Fig. \ref{fig:glbfit_LCDM}, we show that constraints on the parameters of Amati relation from the low-$z$ GRB data in the low-redshift calibration method and from all GRB data in the global fitting method are more or less consistent within $1\sigma$ confidence level. As we know, the constraint on the matter density parameter is strongly correlated with these coefficients of GRB relation. Therefore, the constraint on $\Omega_{\rm m}$ using the global fitting method is quite different from that using the low-redshift calibration method:
\begin{equation}
  \Omega_{\rm m}>0.45
\end{equation}
at $95\%$ confidence level, as shown in the blue line of the upper right panel of Fig. \ref{fig:Omm}. This large difference implies that the obtained constraint on $\Omega_{\rm m}$ using the low-redshift calibration method could be biased and overestimated. On the other hand, the constraint on $\Omega_{\rm m}$ using the global fitting method is far away from the current best-fit values from other observational data $\Omega_{\rm m}\sim0.3$, which might imply that the original Amati relation could not be the best relation to describe the GRB data.

Next, we move to the extended Amati relation. In Table \ref{tab:parameters_LCDM} we also show the constraints on parameters of the extended Amati relation from the low-$z$ GRB data in the low-redshift calibration method and from all GRB data in the global fitting method, which are completely consistent with each other at about $1\sigma$ confidence level. This also implies that the extended Amati relation could ease the tension of constraints on parameters of GRB relation obtained from different GRB data combinations, similar with the conclusion in the previous section \ref{sec:evolution}. Then we compare the constraints on the matter density parameter using these two methods. In Fig. \ref{fig:Omm} we plot the two-dimensional marginalized distribution with 1$\sigma$ and 2$\sigma$ contours between $\Omega_{\rm m}$ and $\alpha$, as well as the one-dimensional distributions of $\Omega_{\rm m}$ and $\alpha$ (blue lines). We also show the constraint on $\Omega_{\rm m}$ using the low-redshift calibration method (red dashed line) in the lower-right panel for comparison.

Clearly, we can see that the constraint on $\Omega_{\rm m}$ using the global fitting method is nearly unconstrained, while the powerful constraint on $\Omega_{\rm m}$ is obtained by using the low-redshift calibration method. After our careful numerical checks, we find that the reason of this huge difference is that the global fitting method includes all the correlations among parameters, such as the matter density parameter and parameters of GRB relation (as shown in the lower-left panel of Fig. \ref{fig:Omm}), during the calculations, while this low-redshift calibration method totally neglects these strong degeneracies by force and only uses the simple error propagation equation [see Eq.(\ref{error})], which only includes the standard deviations of parameters themselves, to estimate the final statistical errors for the data. This will inevitably underestimate the statistical errors of data and then overestimate the constraining power on the matter density parameter. This means that the low-redshift calibration method gives the strongly biased and overestimated constraint on the matter density parameter, which is untrustable.

In the two dimensional contour (lower left panel) of Fig. \ref{fig:Omm}, we add a vertical line which denote the best-fit value $\alpha=0.72$ obtained from the low-redshift GRB data alone in the low-redshift calibration method. This line crosses the $1\sigma$ and $2\sigma$ contours with the crossing values 0.11 $\leq\Omega_{\rm m}\leq$ 0.36 and 0.05 $\leq\Omega_{\rm m}\leq$ 0.76. We find that these values are quite similar with the $1\sigma$ and $2\sigma$, lower and upper limits of the constraint on the matter density parameter by using the low-redshift calibration method (Eq.~\ref{eq:omm:low-z}). This analysis implies that in the low-redshift calibration method, the strong correlations between $\Omega_{\rm m}$ and coefficients of the extended Amati correlation are not fully taken into account in the calculation. Neglecting this information could break the strong degeneracies between $\Omega_{\rm m}$ and correlations coefficients by force and then give quite stringent constraint on the matter density parameter.

Furthermore, in the low-redshift calibration method, we find that the final constraint on $\Omega_{\rm m}$ is very sensitive on how to calibrate the relation. Here, we assume three sets of parameters to calibrate the extended Amati relation: 
\begin{itemize}
  \item (a) the best-fit values obtained from the low-$z$ GRB data;
  \item (b) the best-fit values obtained from all GRB data;
  \item (c) $a=52.40$, $b=1.45$, $\alpha=0.10$ and $\beta=0.14$, which is quite similar with the first case (a), except the $\alpha$.
\end{itemize}
Based on the Table \ref{tab:parameters_LCDM}, we can easily know these three cases are consistent with each other at $1\sigma$ confidence level. Now we use the low-redshift calibration method to constrain $\Omega_{\rm m}$, which is shown in the lower-right panel of Fig. \ref{fig:Omm} [(a) the red dashed line, (b) green dash dot line and (c) black dotted line, respectively]. The case (a) is the standard case in the low-redshift calibration method and obtains the stringent constraint on $\Omega_{\rm m}$ (see Eq.~\ref{eq:omm:low-z}). But if we use case (b) to calibrate the relation, the obtained constraint on the matter density parameter is quite different from that in case (a):
\begin{equation}
  \Omega_{\rm m} = 0.58\pm0.12~(68\%~{\rm C.L.})~.
\end{equation}
The small shift of the parameters of ``calibrated'' relation could significantly change the final constraint on $\Omega_{\rm m}$ in the low-redshift calibration method, which implies that the constraint on $\Omega_{\rm m}$ is very sensitive to the values of parameters for the calibration, when we do not fully take the strong correlations among parameters into account and underestimate the statistical errors. When we use the case (c), the situation becomes worse. The obtained constraint on $\Omega_{\rm m}$ is weaker and only has the lower limit: $\Omega_{\rm m} > 0.71$ at 95\% confidence level.

\begin{figure*}
	\centering
	\includegraphics[width=0.9\textwidth]{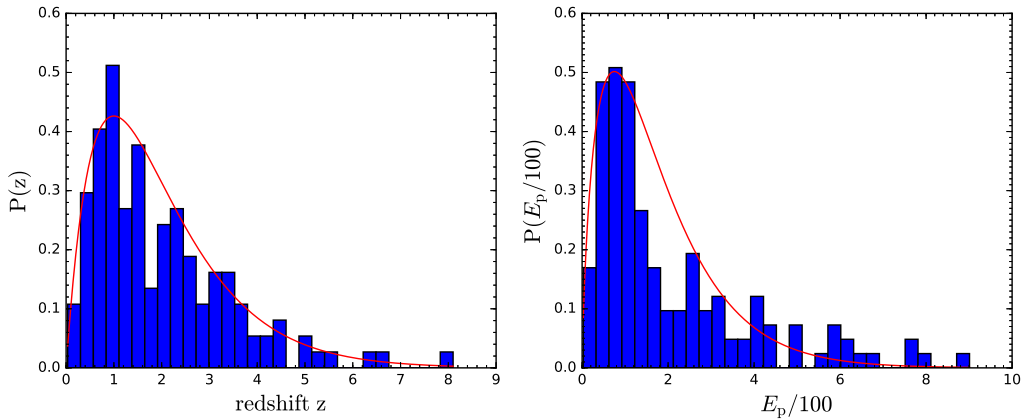}
	\caption{The distributions of redshift $z$ (left panel) and peak energy $E_{\rm p}$ (right panel). The blue histogram represent the distributions of redshift $z$ and peak energy $E_{\rm p}$ respectively. the red lines are the distribution function of this two quantities.}\label{fig:distributions}
\end{figure*}

We also perform the similar check in the original Amati relation. The constraints on the matter density parameter using the cases (a) and (b) to calibrate the relation are shown in the upper-right panel of Fig.\ref{fig:Omm} [(a) the red dashed line and (b) green dash dot line, respectively]. We obtain the same conclusion that the constraint on $\Omega_{\rm m}$ is very sensitive to the values of parameters for the calibration in the low-redshift calibration method.

\subsection{Simulation Results}\label{sec:section3.3}

For the time being, the number of GRB data we can use for possible cosmological studies are very small, when comparing with various type Ia supernovae data sets. There are still very large intrinsic scatter and large statistical errors on data in the current GRB data set, as we discuss above. In order to check how large biased constraint on the matter density parameter by using the low-redshift calibration method, we simulate the mock GRB data with high precisions.

To get the simulation data of GRBs, we first plot the distribution of the redshift $z$ and the observed quantity peak energy $E_{\rm p}$ of the 138 GRBs sample \citep{Liu:2015} in Fig. \ref{fig:distributions} with the blue histogram. Then we find two distribution functions to trace the histogram of redshift $z$ and $E_{\rm p}$. For the redshift $z$, we assumes the distribution function has the form
\begin{equation}\label{equ:distribution_z}
p(z) \propto z e^{-z}.
\end{equation}
We also assumes the distribution of peak energy $E_{\rm p}$ is gamma distribution
\begin{equation}\label{equ:gammadistribution}
p(x;\alpha,\lambda)=\frac{\lambda^{\alpha}}{\Gamma(\alpha)}x^{\alpha-1}e^{-\lambda x},
\end{equation}
where $\alpha$ and $\lambda$ are parameters, and the gamma function is
\begin{equation}
\Gamma (\alpha) = \int_{0}^{\infty} e^{-t}t^{\alpha-1}dt.
\end{equation}
We put the distribution function of $E_{\rm p}$ in Fig. \ref{fig:distributions} (right panel) the form of
\begin{equation}\label{equ:distribution_Ep}
p(E_{\rm p}/100) \propto \frac{1}{\Gamma(\alpha)}(E_{\rm p}/100)^{\lambda}e^{-E_{\rm p}/100}.
\end{equation}
We also use the distribution of $E_{\rm p,i}$ to generate the mock GRB data, and find that the obtained constraints on the parameters of relation and the matter density parameter are consistent with that by using the distribution of $E_{\rm p}$.

We use the fiducial values of the extended Amati relation: $a=52.52$, $b= 1.59$, $\alpha= 0.33$, $\beta= -0.31$, obtained from all-redshift GRB data by using the global fitting calibration method in the flat $\Lambda$CDM framework. (Different choices of fiducial values do not affect the final results.) Based on these two distribution functions, equations (\ref{equ:distribution_z}) and (\ref{equ:distribution_Ep}), we could obtain the simulation data for both redshift $z$ and peak energy $E_{\rm p}$, and then get the simulation data of bolometric fluence $S_{\rm bolo}$ according to Eq.~(\ref{equ:newrelation}). For the real GRB data, we find that the uncertainties of $E_{\rm p}$ and $S_{\rm bolo}$ have the mean relative error of $28\%$ and $18\%$, respectively \citep{Liu:2015}. Therefore, here we set the simulation data of $E_{\rm p}$ and $S_{\rm bolo}$ have the same relative errors. Furthermore, in the process of simulation, we also include the information of the intrinsic scatter and choose $\sigma_{\rm int}=0.353$ obtained from the calibration method using all current GRB data. Finally, the $\chi^2$ equation becomes
\begin{equation}\label{equ:sim_chi2}
\chi^{2}(a,b,\alpha,\beta)=\sum_{i}\frac{\left(y_{i}-A-Bx_{i}\right)^{2}}{\sigma^{\prime 2}_{y_{i}}+B^{2}\sigma_{x_{i}}^{2}}.
\end{equation}
where $\sigma^{\prime}_{y_{i}}=\sqrt{\sigma^{2}_{\rm int}+\sigma^{2}_{y_{i}}}$ which includes the intrinsic scatter.

Firstly, we use the current precision ($\sigma_{E_{\rm p}}$,$\sigma_{S_{\rm bolo}}^{\prime}$) to simulate the 138 GRB data to constrain the coefficients of the extended Amati relation and obtain the similar best-fit values and error bars of coefficients with those from the current real GRB data, which means that we could safely use this mock data for further calculations.

\begin{figure}
	\centering
	\includegraphics[width=0.45\textwidth]{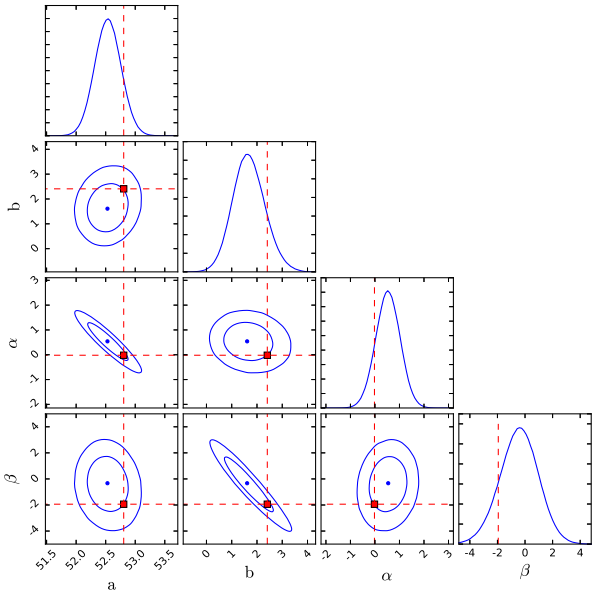}\\
	\caption{The 1-D and 2-D marginalized distributions with $1\sigma$ and $2\sigma$ contours for the coefficients $a$, $b$, $\alpha$ and $\beta$ from the mock GRB data with the current precision of GRB data.}\label{fig:1sigma_point_1}
\end{figure}

Next, we use the low-redshift calibration method to constrain the matter density parameter from this mock GRB data. In practice, we select the low-redshift GRB samples with $z<1.4$ to constrain those four coefficients of the extended Amati relation. The results are shown in Fig. \ref{fig:1sigma_point_1}. The obtained best-fit values perfectly recover the fiducial ones of parameters of GRB relation. In section \ref{sec:section3.2}, we have proven that a small shift of the parameters of ``calibrated'' relation could significantly change the final constraint on $\Omega_{\rm m}$ from the real GRB data in the low-redshift calibration method. Therefore, here we want to check this using the mock GRB data. We assume the values of parameters (52.805, 2.408, -0.020, -1.939), which is a small shift from the best fit fiducial values and safely lay in the $1\sigma$ contour (red points in Fig. \ref{fig:1sigma_point_1}), to calibrate the extended Amati relation. Then we use the mock high-redshift GRB data to constrain the matter density parameter, based on this ``calibrated'' relation, and obtain the 68\% C.L. constraint:
\begin{equation}
\Omega_{\rm m}=0.65\pm0.15~,
\end{equation}
which is far away from the fiducial value $\Omega_{\rm m}=0.308$. This result tells us that even a small shift of coefficients in the $1\sigma$ contour, the constraint on the matter density parameter with the current precision of GRB data could still be significantly biased using the low-redshift calibration method. The constraint on $\Omega_{\rm m}$ is very sensitive to the values of parameters for the calibration, when we do not fully take the strong correlations among parameters into account and underestimate the statistical errors in the low-redshift calibration method.

\begin{figure}
	\centering
	\includegraphics[width=0.45\textwidth]{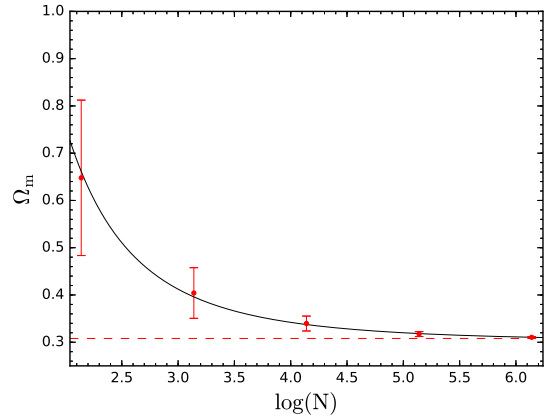}\\
	\caption{The best-fit value and 1$\sigma$ error of $\Omega_{\rm m}$ for different quantities of GRBs. The red points with error bars represent the best-fit value and 1$\sigma$ error of $\Omega_{\rm m}$ for different numbers of GRBs respectively. The red dashed line corresponding to the fiducial value of $\Omega_{\rm m}=0.308$.}\label{fig:find_min_grb_all}
\end{figure}

In Fig \ref{fig:find_min_grb_all} we show the obtained biased values of matter density parameter from the mock GRB data with different precisions using the low-redshift calibration method. We can see that when GRB data have larger samples with the current precision or the relative errors of GRB data decreases, $\Omega_{\rm m}$ will be more and more close to the fiducial value.

\section{Conclusions and discussions}\label{sec:summary}

Thanks to the extremely high power of the explosion, Gamma-ray bursts (GRBs) are proposed as a promising candidate to trace the Hubble diagram of the Universe in high redshift range. In recent years, several popular luminosity correlations have been statistically concluded from GRBs observations. People also have made great efforts to standardize them as cosmological distance indicators. In general, there are two key issues extensively discussed in the literature when we use these correlations for cosmology. The first issue is that whether luminosity correlations evolve with redshift. The other one is about methods to calibrate luminosity correlations. In this paper we take the Amati relation as an example and use the current GRB data in the redshift range [0.0331,~8.1] to investigate these two issues. Here we summarize our main conclusions in more detail:
\begin{itemize}
  \item We divide the whole GRB data into two redshift bins and constrain the coefficients $a$ and $b$ of the original Amati relation in each bin, respectively. Based on the MCMC method, we find that the constraints on both the intercept $a$ and the slope $b$ are quite different from the low-redshift and high-redshift GRB data, respectively. The tension is more than $1\sigma$ level, which implies that the parameters of original Amati relation could be redshift-dependent.
  \item We introduce two extra redshift-dependent terms to characterize the redshift evolution of the luminosity correlation. Interestingly, we find that the tension between constraints of parameters from low-$z$ and high-$z$ GRB data are alleviated.
  \item Besides the evolution of Amati correlation, the calibration is also very important for using GRB data in cosmological studies. We firstly check the constraint on the matter density parameter using the low-redshift calibration method and obtain $\Omega_{\rm m}=0.24\pm0.12$ (68\% C.L.) which is consistent with other works.
  \item However, we could also use the global fitting method to calibrate the GRB data, in which we use all the GRB data to constrain coefficients of the extended Amati relation and the cosmological parameters simultaneously. In this case, we find that, due to the current poor precision of the GRB data, the constraint on $\Omega_{\rm m}$ is very weak, which is quite different from that by using the low-redshift calibration method.
  \item After our careful checks, we find that the low-redshift calibration method does not take the whole correlations between $\Omega_{\rm m}$ and coefficients into account. Neglecting the correlation information can break the degeneracies between $\Omega_{\rm m}$ and coefficients. Therefore, the obtained constraint on $\Omega_{\rm m}$ is totally biased. A small shift of the parameters of ``calibrated'' relation could significantly change the final constraint on $\Omega_{\rm m}$ in the low-redshift calibration method, which implies that the constraint on $\Omega_{\rm m}$ is very sensitive to the values of parameters for the calibration.
  \item In order to investigate how large biased constraint on $\Omega_{\rm m}$ by using the low-redshift calibration, we simulate the mock GRB data with different precisions. We find that the mock data with current precision will give significantly biased result using the low-redshift calibration method, even we assume a small shift on the parameters of ``calibrated'' relation. When GRB data include larger samples with the current precision or the relative errors of GRB data decreases, the constraint on $\Omega_{\rm m}$ is more and more close to the fiducial value.
\end{itemize}

\section*{Acknowledgements}
G.-J. Wang and Z.-H. Zhu are supported by the the National Science Foundation of China under grant No. 11373014. H. Yu is supported by the National Basic Research Program of China (973 Program, grant No. 2014CB845800). Z.-X. Li is supported by the NSFC under grant No. 11505008. J.-Q. Xia is supported by the National Youth Thousand Talents Program and the NSFC under grant No. 11422323. The research is also supported by the Strategic Priority Research Program ``The Emergence of Cosmological Structures'' of the Chinese Academy of Sciences, Grant No.XDB09000000, and the NSFC under grant Nos. 11633001 and 11690023.

\end{document}